\documentstyle[11pt,newpasp,twoside,epsf]{article}

\def\edcomment#1{\iffalse\marginpar{\raggedright\sl#1\/}\else\relax\fi}
\reversemarginpar

\newcommand {\hi} {H\,{\small I}}
\newcommand {\km} {km s$^{-1}$}
\newcommand {\ci}{$^{\circ}$}
\newcommand {\mo}{$M_{\odot}$}

\begin{document}
\title{High Velocity Gas in the Halos of Spiral Galaxies}
\author{Filippo Fraternali \& Tom Oosterloo}
\affil{ASTRON, Dwingeloo, NL}
\author{Rense Boomsma}
\affil{Kapteyn Institute, Groningen, NL}
\author{Rob Swaters}
\affil{Johns Hopkins University, Baltimore, USA}
\author{Renzo Sancisi}
\affil{INAF, Osserv. Astron. Bologna, I \& Kapteyn Institute, Groningen, NL}

\begin{abstract}
Recent, high sensitivity, \hi\ observations of nearby spiral galaxies show that their thin `cold' disks are surrounded by thick layers (halos) of neutral gas with 
anomalous kinematics.
We present results for three galaxies viewed at different inclination angles:
NGC\,891 (edge-on), NGC\,2403 (i$=$60\ci), and NGC\,6946 (almost face-on).
These studies show the presence of halo gas up to distances of 10$-$15 kpc from the plane.
Such gas has a mean rotation 25$-$50 \km\ lower than that of the gas in the plane, and
some complexes are detected at very high velocities, up to 200$-$300 \km.
The nature and origin of this halo gas are poorly understood. 
It can either be the result of a galactic fountain or of accretion from the intergalactic medium.
It is probably the analogous of some of the High Velocity Clouds (HVCs) of the Milky Way.
\end{abstract}

\section{Introduction}

In recent years, deep \hi\ surveys of nearby spiral galaxies, viewed at different inclination angles, have been obtained to study the gas outside the plane of the disk.
Edge-on galaxies have been used to investigate the vertical distribution and rotation velocity and face-on galaxies to study the distribution in the plane and the vertical motions.

The pioneering study of the edge-on galaxy NGC\,891 (Swaters, Sancisi \& Van der Hulst 1997) revealed an \hi\ halo ($\sim$15\% of the total \hi\ mass) extended up to 5 kpc from the plane and having a mean rotation velocity 25$-$100 \km\ lower than that of the disk.
This velocity gradient has also been observed in the ionized gas (e.g.\ NGC\,5775; Rand 2000).
More recently, detection of extra-planar \hi\ has been reported for other edge-on systems such as the superthin galaxy UGC\,7321 where there is also the indication of a rotation velocity gradient (Matthews \& Wood 2003).

Several studies of face-on galaxies have shown the presence of high
velocity gas complexes and of 
holes in the \hi\ distribution (e.g.\ Ho\,{\small II}, Puche 1992).
In some cases the high velocity gas is 
associated with the \hi\ holes (e.g.\ M101, Kamphuis, Sancisi \& Van der
Hulst 1991) and thus possibly caused by star formation activity. 
In others, it is probably due to accretion of material from the
intergalactic medium (e.g.\ Van der Hulst \& Sancisi 1988).

\section{The spiral galaxy NGC\,2403}

NGC\,2403 is a non-interacting (Sc) spiral galaxy, very similar to M\,33 and located at a distance of about 3 Mpc.
Deep \hi\ observations with the VLA (Fraternali et al.\ 2002a) have revealed the presence of a gas component with anomalous kinematics, not expected for a thin disk (Figure\,1).
The anomalous \hi\ in NGC\,2403 forms a thick (scale-height$\sim$1$-$3 kpc) layer with a mass of 3 $\times$ 10$^8$ $M_{\odot}$ (1/10 of the total \hi\ mass) and a mean rotation velocity 20$-$50 km~s$^{-1}$ lower than that of the gas in the disk.
It is the analogous of the \hi\ halo found by Swaters et al.\ (1997) in NGC\,891.
Part of the anomalous gas, located in the inner parts of the galaxy, shows large non-circular motions with projected differences of up to 150 \km\ from rotation (300 \km\ if it is moving vertically).
High velocity complexes are also observed outside the bright optical disk, the most remarkable being an 8 kpc long filament with a mass of about 10$^7$ \mo.
The halo gas in NGC\,2403 also shows the indication of a radial infall (of 10$-$20 km~s$^{-1}$) towards the centre of the galaxy (Fraternali et al.\ 2001).

\begin{figure}[ht]
\plottwo{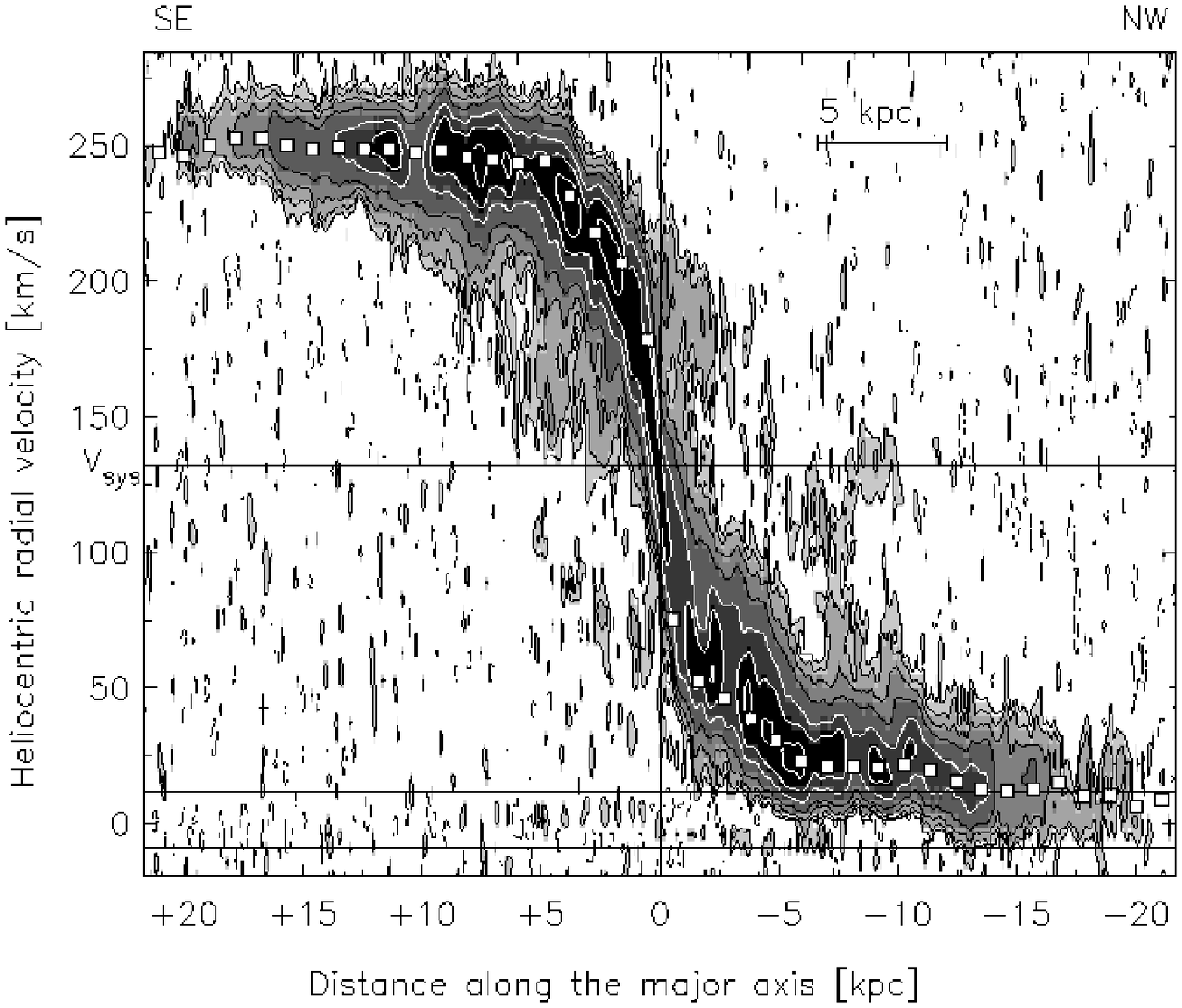}{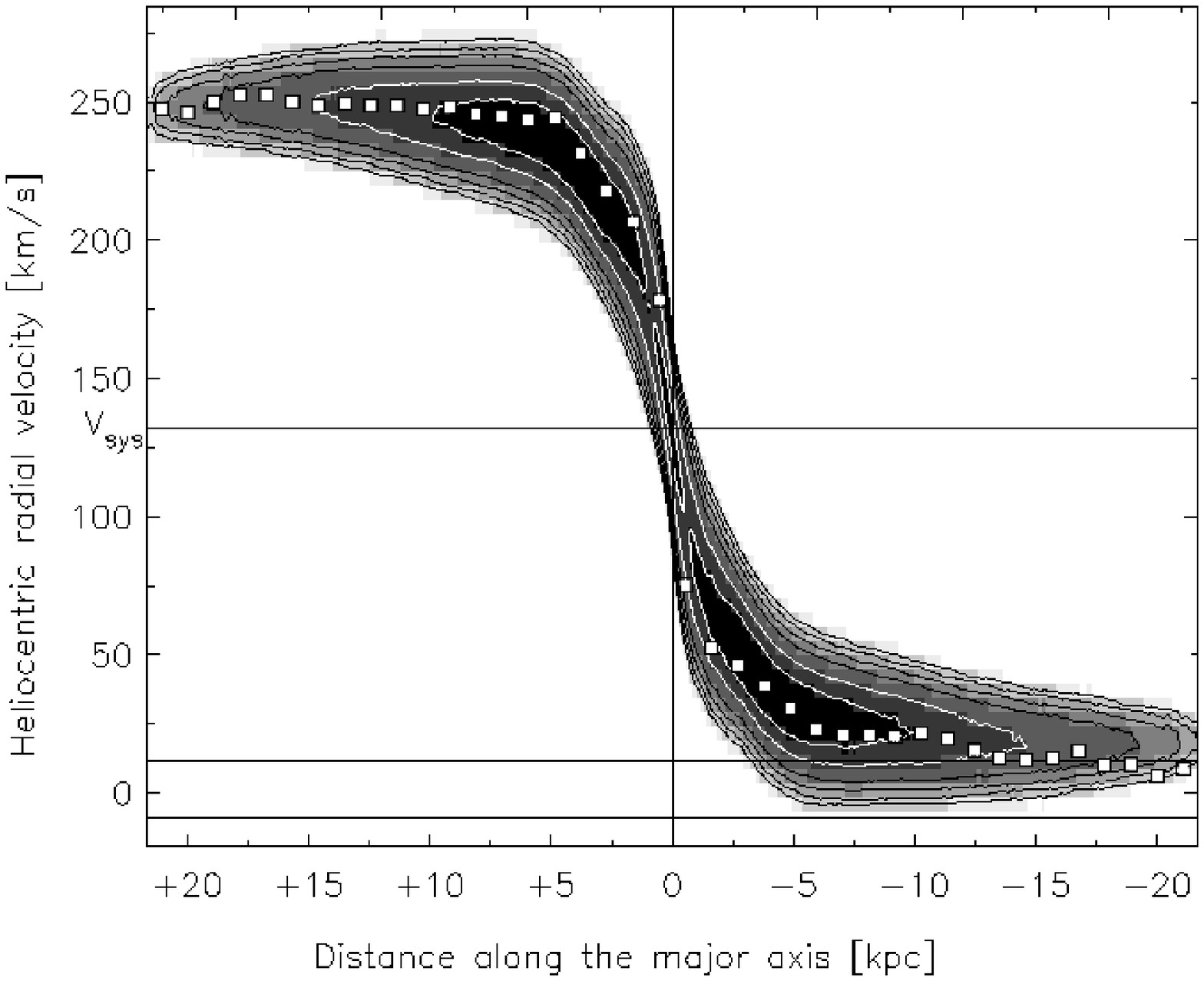}
\caption{Left panel: \hi\ position-velocity diagram along the major axis of NGC\,2403.
The white squares show the rotation curve.
Right panel: model for a thin \hi\ disk with Gaussian velocity
dispersion.
The \hi\ not-reproduced by the model is the anomalous (halo) gas.}
\end{figure}

The origin of the anomalous gas in NGC\,2403 is uncertain. 
It may be explained by a galactic fountain mechanism (Shapiro \& Field 1976). 
The discovery of hot, diffuse, X-ray emitting gas with $Chandra$ (Fraternali et al.\ 2002b) seems to support this interpretation.
However an external (accretion) origin can not be excluded.

\section{The edge-on galaxy NGC\,891}

NGC\,891 is a nearby (Sbc) spiral galaxy seen almost perfectly edge-on
(i$\geq$88.6\ci; Rupen et al.\ 1987).
It is located at a distance of 9.5  Mpc and it is considered very
similar to the Milky Way. 
A recent \hi\ study of this galaxy has shown the presence of an extended \hi\ halo rotating more slowly (25$-$100 \km) than the disk (Swaters et al.\ 1997).

We have re-observed NGC\,891 with the Westerbork Synthesis Radio
Telescope (WSRT) with about 200 hrs of integration (Oosterloo et al.,
in preparation).
The high sensitivity and long integration of these new observations have permitted the detection of \hi\ gas at distances of about 10$-$15 kpc from the plane.
Figure 2 shows a comparison between the total \hi\ map of NGC\,891 published by Swaters et al.\ (1997) and the one produced with the new WSRT data.
Our new observations are about a factor 5 better in sensitivity and
reveal an \hi\ layer much more extended in the vertical direction (about $\times$2) while the on-plane size is almost unchanged.

\begin{figure}[ht]
\plotone{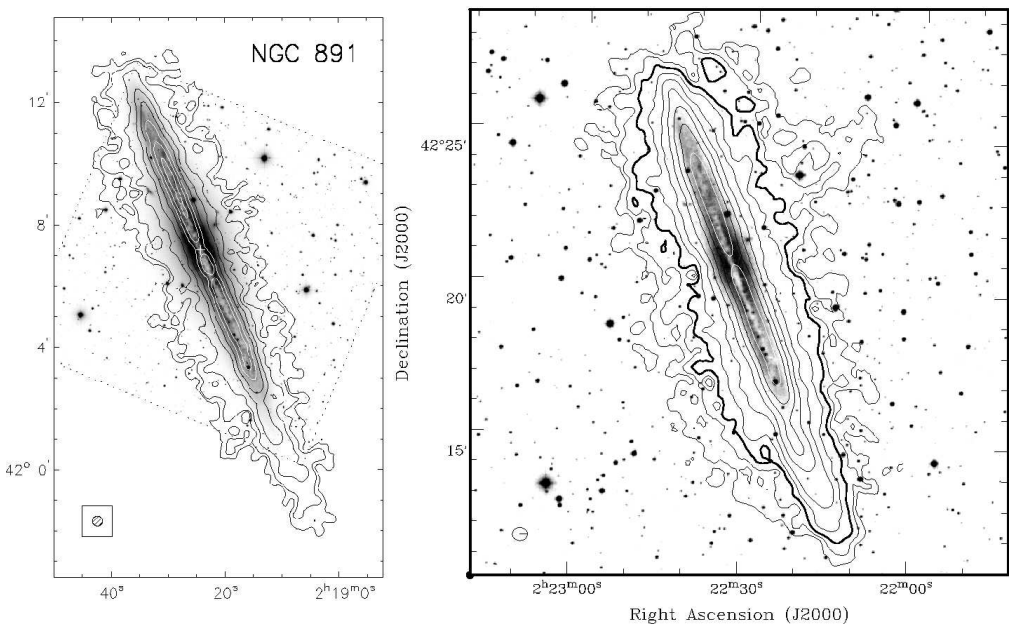}
\caption{A comparison between the total \hi\ map published by Swaters et
al.\ 1997 (left) and the one obtained with our new WSRT data
(right). Contours in the right map are: 1.7, 4.5, 9, 18.5, 37,
74, 148, 296.5, 593 $\times$ 10$^{19}$ atoms cm$^{-2}$.
The thicker contour (third from the last) roughly reproduces the lowest contour in the left map.
}
\end{figure}

\begin{figure}[ht]
\plotone{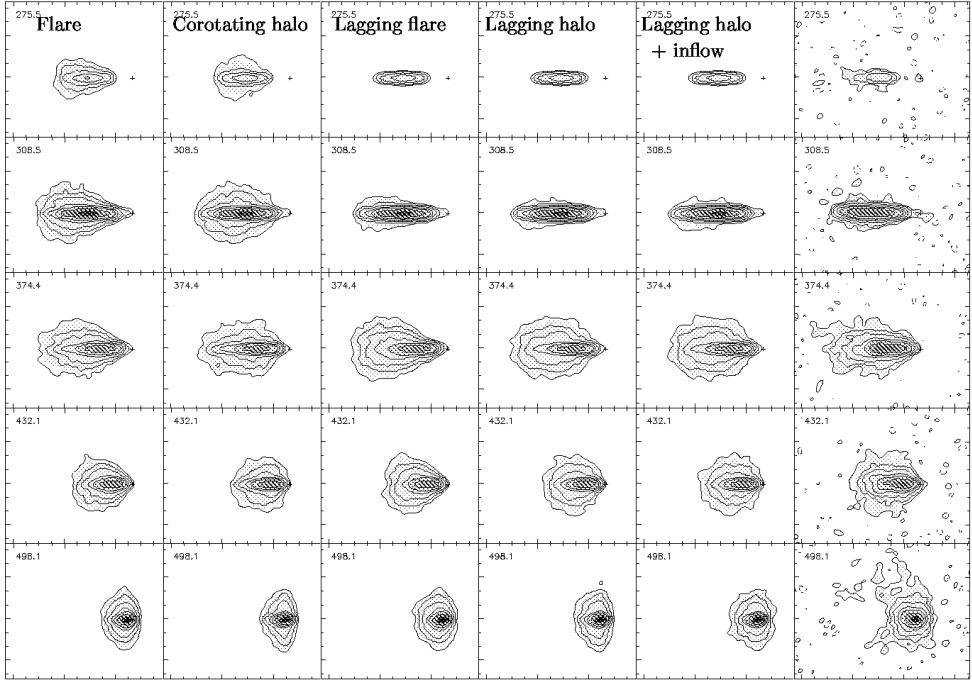}
\caption{Channel maps for NGC\,891 compared with models (see text).}
\end{figure}

Swaters et al.\ have modelled the \hi\ layer of NGC\,891 and excluded the possibility that the extra-planar gas could be produced by a warp along the line of sight or an imperfectly edge-on orientation of the system.
Our modeling analysis confirms these results. 
Moreover we have made a more detailed study of the halo gas.
In Figure 3 we show the observed channels maps of NGC\,891 (right hand column) compared with several models of the \hi\ layer.
The flare model (column 1) shows the effect of an increase of the \hi\ scale-height (up to FWHM$\sim$6 kpc) in the outer disk.
The corotating halo model (column 2) is a 2-component model formed by a thin disk plus a thicker disk (FWHM=6 kpc), comprising about 20\% of the total \hi\ mass, with the same rotation curve as the disk.
These two models fail to reproduce the thin channels at high rotation
velocities (275.5 and 308.5 \km). 
The other three models are made with 2 components: disk and lagging halo. 
The halo has a constant rotation velocity of 185 \km (about 35 \km\ lower than the disk).
In the lagging flare, the extra-planar gas is mainly located in the outer parts of the disk, and in the other two models it has a distribution (in R) similar to that of the gas in the thin disk.
In the last model (column 5) the halo gas also has a radial infall motion (of 20 \km), similar to that found in NGC\,2403. 
All the models with lagging halos are acceptable with some preference for models in column 4 and 5.

From this analysis we can conclude that the extra-planar \hi\ in
NGC\,891 is rotating more slowly (mean velocity of about 35 \km) than the gas in the plane.
It is distributed in a thick (FWHM$\sim$6 kpc) layer with a radial profile similar to that of the disk.
Moreover, the data are consistent with a radial inflow of halo gas toward the centre of the galaxy.

\section{The nearly face-on galaxy NGC\,6946}

NGC\,6946 is a nearby Scd spiral galaxy, viewed at an inclination angle i$\simeq$30\ci\, and located at a distance of 5.9 Mpc.
It has been studied in \hi\ by Kamphuis (1993) and this revealed an extended \hi\ disk (about twice the size of the optical) and the presence of several \hi\ holes.

NGC\,6946 has recently been re-observed in \hi\ with the WSRT (Boomsma
et al., in preparation). 
Figure 4 shows the total \hi\ map obtained from these new data, the
ellipse outlines the area inside R$_{25}$=5.5$'$.  
The \hi\ disk is extended far beyond the optical radius out to
$\sim$13$'$$\simeq$22 kpc. 
The \hi\ distribution clearly shows a spiral pattern and numerous
holes, mostly confined to the inner (bright optical) part of the
disk. 

\begin{figure}[ht]
\plottwo{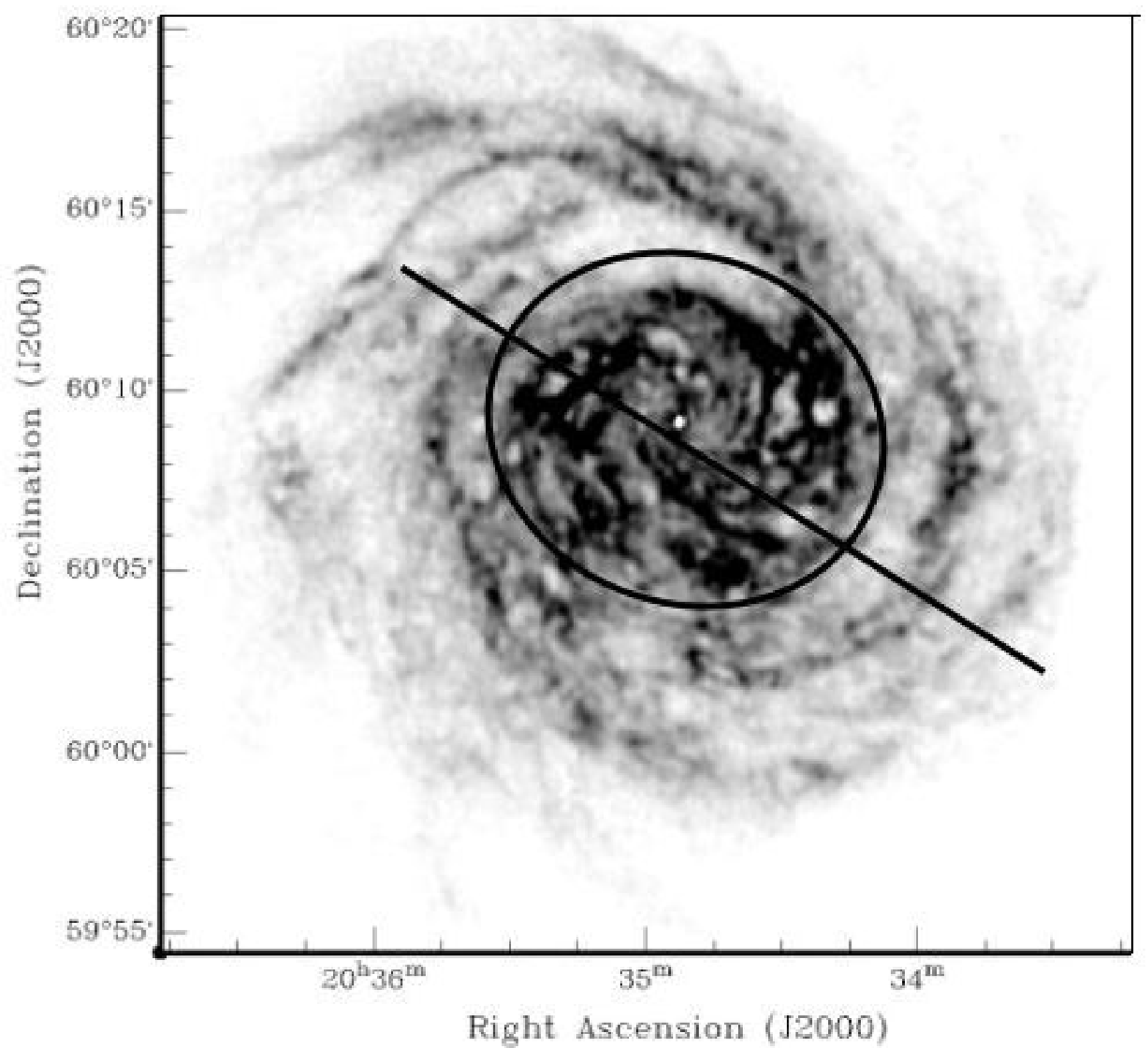}{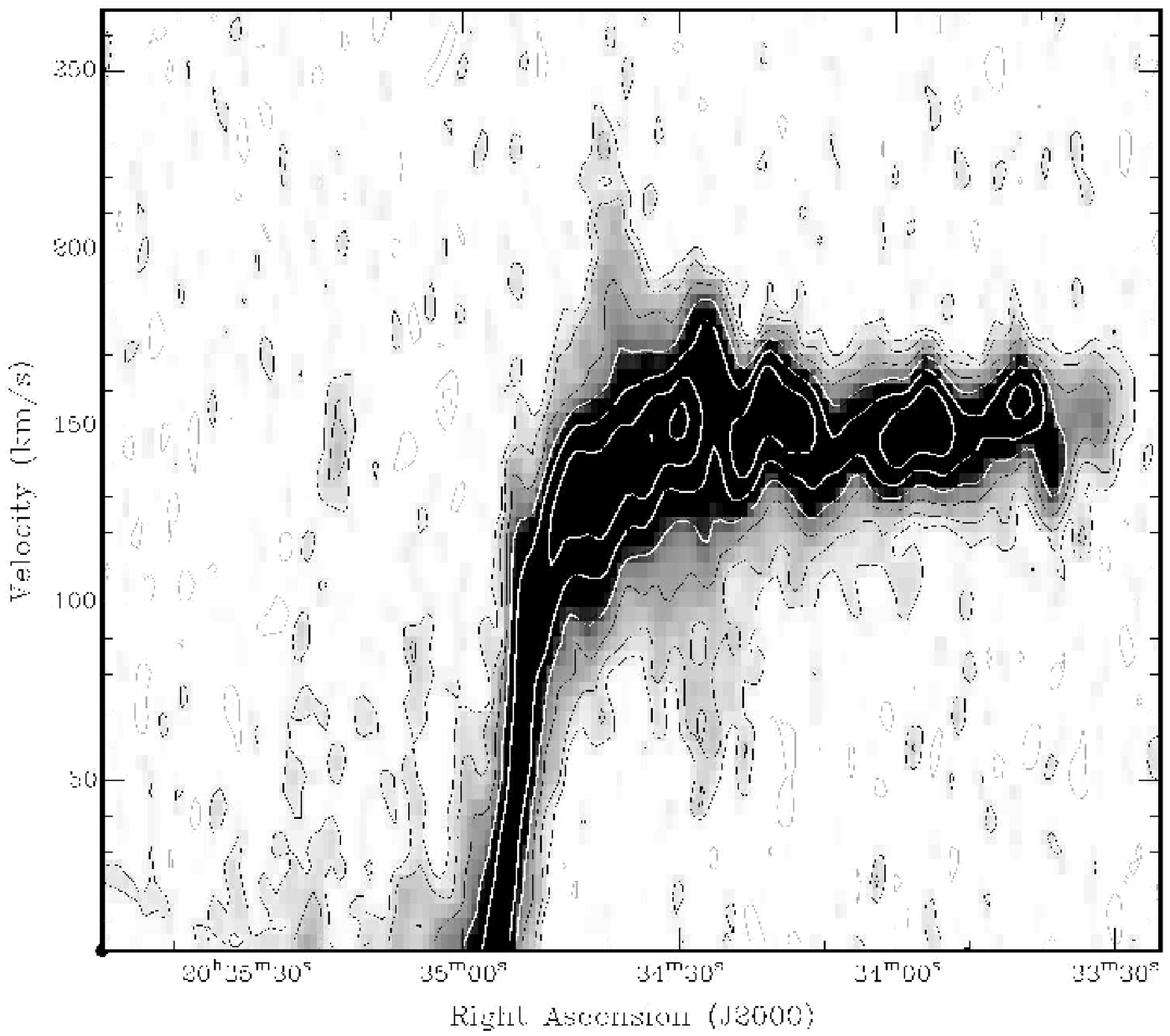}
\caption{Left panel: total \hi\ map for NGC\,6946 obtained with the WSRT. 
The ellipse indicates the $size$ of the optical disk (R$_{25}$).
Note the numerous holes in the \hi\ distribution.
Right panel: position-velocity diagram along the line overlaid on the map in the left panel. 
}
\end{figure}

High velocity gas is detected all over the disk of NGC\,6946, in some
cases associated with \hi\ holes. 
In Figure 4 (right panel) we show a position-velocity diagram taken
roughly along the major axis (indicated by the line on
the total \hi\ map).
This diagram illustrates the variety of features which are detected in NGC\,6946.
There is clearly anomalous (halo) gas similar to that found in NGC\,2403. 
However, in this case, part of the anomalous gas is also observed on the high velocity side of the diagram (showing the presence of vertical motions).
Moreover, in NGC\,6946 clouds at very high velocity are observed (see the one at projected velocity of 150 \km).
Surprisingly, some of these are detected outside the bright optical disk (Boomsma et al., this conference) and seem therefore not to be related with star formation.

\section{Discussion and conclusions}

The study of spiral galaxies viewed at different inclination angles is
needed to reconstruct the 3D distribution and kinematics of the halo
gas.  
The three galaxies shown here are good complementary candidates for such a study.
What we learn about the halo gas from these three studies can be summarized as follows:
\begin{enumerate}
\item{{\bf Distribution:} 
Spiral galaxies have thick components (halos) of \hi\ gas with
typical FWHMs of a few kpc. \hi\ clouds are detected up to distances of
10$-$15 kpc from the plane of the disk (NGC\,891).
The typical mass of the \hi\ halo is about 10$-$20\% of the total
\hi\ mass of the galaxy.
The typical masses of individual clouds are a few 10$^6$ \mo, very similar to 
some of the HVCs in the Milky Way (e.g.\ complex A, Van Woerden et al.\ 1999).
}
\item{{\bf Overall kinematics:}
The halo gas is rotating more slowly (20$-$50 \km) than the gas in the
plane.
The difference in rotation velocity seems to increase in the central
regions (NGC\,2403).
The halo gas in NGC\,2403 also shows a radial infall towards the centre.
}
\item{{\bf High Velocity Clouds:} 
Several gas components with high (vertical?) velocities have been found (NGC\,2403,
NGC\,6946). 
The difference between their velocities and the regular rotation velocity reaches values of 200$-$300 \km. 
The largest differences are observed in the inner regions (NGC\,2403). 
In some cases (not always) they are associated with holes in the \hi\ distribution.
}
\end{enumerate}

The origin and nature of the halo gas remain uncertain.
The rotation velocity gradient, the diffuse hot gas found in NGC\,2403, and the concentration of the high velocity clouds in the central (star forming) part seem to indicate that it is produced by a galactic fountain process (e.g.\ Bregman 1980).
Other facts such as the presence of massive, filamentary complexes (NGC\,2403; M\,33, J.M.\ van der Hulst, private communication) and the presence of high velocity clouds beyond the optical disk (NGC\,6946, NGC\,2403) have no obvious explanations in a galactic fountain model and may point at an external origin.

\end{document}